\documentclass[apj,iop]{emulateapj}

\usepackage[colorlinks=true,citecolor=blue]{hyperref}
\usepackage{amssymb}
\usepackage{subfigure}
\usepackage{longtable}
\usepackage{tablefootnote}
\usepackage{savesym}
\savesymbol{tablenum}
\usepackage{siunitx}
\restoresymbol{SIX}{tablenum}
\let\oldenddeluxetable\enddeluxetable
\let\olddeluxetable\deluxetable
\makeatletter
\renewenvironment{deluxetable}[1]{
\olddeluxetable{[#1]}
\def\pt@format{#1}%
}{\oldenddeluxetable}
\makeatother

\shorttitle{}
\shortauthors{Strait et al.}

\begin{document}

\title{Stellar Properties of $\MakeLowercase{z}  \gtrsim 8$ Galaxies in the Reionization Lensing Cluster Survey}

\author{Victoria Strait\altaffilmark{1}}
\author{Maru{\v s}a Brada{\v c}\altaffilmark{1}}
\author{Dan Coe\altaffilmark{2}}
\author{Larry Bradley\altaffilmark{2}}
\author{Brett Salmon\altaffilmark{2}}
\author{Brian C. Lemaux\altaffilmark{1}}
\author{Kuang-Han Huang\altaffilmark{1}}
\author{Adi Zitrin\altaffilmark{3}}
\author{Keren Sharon\altaffilmark{4}}
\author{Ana Acebron\altaffilmark{3}}
\author{Felipe Andrade-Santos\altaffilmark{5}}
\author{Roberto J. Avila\altaffilmark{2}}
\author{Brenda L. Frye\altaffilmark{6}}
\author{Austin Hoag\altaffilmark{7}}
\author{Guillaume Mahler\altaffilmark{4}}
\author{Mario Nonino\altaffilmark{8}}
\author{Sara Ogaz\altaffilmark{2}}
\author{Masamune Oguri\altaffilmark{9,10,11}}
\author{Masami Ouchi\altaffilmark{11,12}}
\author{Rachel Paterno-Mahler\altaffilmark{13}}
\author{Debora Pelliccia\altaffilmark{1,14}}
%\author{Avery Peterson\altaffilmark{4}} 

%\author{Daniel P. Stark\altaffilmark{9}} % 65 spectroscopy
%\author{Ramesh Mainali\altaffilmark{9}} % 19 spectroscopy
%\author{Pascal A. Oesch\altaffilmark{10}} % 41 high-z proposals
%\author{Michele Trenti\altaffilmark{11,12}} % 78
%\author{Daniela Carrasco\altaffilmark{11}} % [0000-0002-3772-0330] 25 completeness %simulations w/ Michele
%\author{William A. Dawson\altaffilmark{13}} % [0000-0003-0248-6123]48+171 ground-based %imaging; faint arc detection
%\author{Christine Jones\altaffilmark{14}} % 143
 % [0000-0003-1625-8009]162
%\author{Nicole G. Czakon\altaffilmark{15}} % 136
%\author{Keiichi Umetsu\altaffilmark{15}} % [0000-0002-7196-4822] 150
%\author{Benedetta Vulcani\altaffilmark{16}} % [0000-0003-0980-1499] 16 + comments

\affil{\altaffilmark{1}Physics Department, University of California, Davis, CA 95616, USA}
\affil{\altaffilmark{2}Space Telescope Science Institute, Baltimore, MD 21218, USA}
\affil{\altaffilmark{3}Department of Physics, Ben-Gurion University, Be’er-Sheva 84105, Israel}
\affil{\altaffilmark{4}Department of Astronomy, University of Michigan, 1085 South University Ave, Ann Arbor, MI 48109, USA}
\affil{\altaffilmark{5}Harvard-Smithsonian Center for Astrophysics, 60 Garden Street, Cambridge, MA 02138, USA}
\affil{\altaffilmark{6}Department of Astronomy, Steward Observatory, University of Arizona, 933 North Cherry Avenue, Tucson, AZ, 85721, USA}
\affil{\altaffilmark{7}Department of Physics and Astronomy, University of California, Los Angeles, CA 90095-1547, USA}

\affil{\altaffilmark{8}INAF -- Osservatorio Astronomico di Trieste, via G. B. Tiepolo 11, I-34131
Trieste, Italy}
\affil{\altaffilmark{9}Research Center for the Early Universe, University of Tokyo, 7-3-1 
Hongo, Bunkyo-ku, Tokyo 113-0033, Japan}
\affil{\altaffilmark{10}Department of Physics, The University of Tokyo, 7-3-1 Hongo, Bunkyo-ku, Tokyo 113-0033, Japan}
\affil{\altaffilmark{11}Kavli Institute for the Physics and Mathematics of the Universe (Kavli IPMU, WPI), University of Tokyo, Kashiwa, Chiba 277-8583, Japan}
\affil{\altaffilmark{12}Institute for Cosmic Ray Research, The University of Tokyo, 5-1-5 Kashiwanoha, Kashiwa, Chiba 277-8582, Japan}
%\affil{\altaffilmark{7}Astronomy Department and Institute for Astrophysical Research, Boston University, 725 Commonwealth Ave., Boston, MA 02215, USA}
\affil{\altaffilmark{13}WM Keck Science Center, 925 N. Mills Avenue, Claremont, CA 91711}
\affil{\altaffilmark{14}Department of Physics and Astronomy, University of California, Riverside, CA 92521, USA}

\begin{abstract}
Measurements of stellar properties of galaxies when the universe was less than one billion years old yield some of the only observational constraints of the onset of star formation. We present here the inclusion of \textit{Spitzer}/IRAC imaging in the spectral energy distribution fitting of the seven highest-redshift galaxy candidates selected from the \emph{Hubble Space Telescope} imaging of the Reionization Lensing Cluster Survey (RELICS). We find that for 6/8 \textit{HST}-selected $z\gtrsim8$ sources, the $z\gtrsim8$ solutions are still strongly preferred over $z\sim$1-2 solutions after the inclusion of \textit{Spitzer} fluxes, and two prefer a $z\sim 7$ solution, which we defer to a later analysis. We find a wide range of intrinsic stellar masses ($5\times10^6 M_{\odot}$ -- $4\times10^9$ $M_{\odot}$), star formation rates (0.2-14 $M_{\odot}\rm yr^{-1}$), and ages (30-600 Myr) among our sample. Of particular interest is Abell1763-1434, which shows evidence of an evolved stellar population at $z\sim8$, implying its first generation of star formation occurred just $< 100$ Myr after the Big Bang. 
SPT0615-JD, a spatially resolved $z\sim10$ candidate, remains at its high redshift, supported by deep \textit{Spitzer}/IRAC data, and also shows some evidence for an evolved stellar population. 
Even with the lensed, bright apparent magnitudes of these $z \gtrsim 8$ candidates (H = 26.1-27.8 AB mag), only the \textit{James Webb Space Telescope} will be able further confirm the presence of evolved stellar populations early in the universe.
\end{abstract}

\keywords{}

\section{Introduction}
High-$z$ galaxies are key sources in the epoch of reionization, and to understand the contributions of the faint \mbox{$z\sim8-10$} population by way of ionizing photon production, we need measurements of star formation rate (SFR) and stellar mass. However in practice, robust constraints on physical properties of \mbox{$z \sim 8-10$} galaxies are difficult to place. Surveys using lensing %, e.g., CLASH \citep{post12}, HFF \citep{lotz17} and BoRG/HIPPIES \citep{trenti11,yan11} 
and blank fields %, e.g., GOODS \citep{dickinson2003}, CANDELS \citep{grogin11,koekemoer11} and XDF \citep{illingworth2013}, 
to target high-$z$ galaxies in recent years have rapidly grown the sample. In particular, measurements of ages of galaxies in the high-$z$ universe have provided one of the few observational probes of the onset of star formation (e.g., \citealp{egami05,richard11,hua16}). The most recent spectroscopically confirmed example by \cite{hashimoto18} (see also \citealp{zheng12,bradac14,hoag18}) implies first star formation at $\sim$ 250 Myr after the Big Bang as evidenced by an old stellar population in the galaxy MACS1149-JD.

There are also a number of galaxies that are not yet spectroscopically confirmed and show signs of a possible evolved stellar population at high-$z$. At \mbox{$z\sim8$}, spectral energy distribution (SED) results are heavily influenced by near-IR fluxes, since the Balmer/$D_{n}$(4000) break (hereafter Balmer break) falls into \textit{Spitzer} channel 1 (3.6$\mu$m, [3.6] or ch1 hereafter) from $z\sim7-10$, requiring \textit{Spitzer} fluxes for robust measurements of stellar mass, SFR, and age. Complicating the problem, strengths of nebular emission lines and dust content at these redshifts are unknown, creating a degeneracy between emission lines and the Balmer break that is difficult to disentangle with the currently available near-IR broadband observations. When a spectroscopic redshift is available, it is sometimes possible to disentangle the degeneracy if the emission lines fall outside of a broadband, as in \cite{hashimoto18}. %The equivalent width distribution of these emission lines is important is removing bias from stellar properties \citep{labbe13}. 
While the \textit{James Webb Space Telescope} (JWST) will ultimately be able to break most of these degeneracies, identifying candidates with broadband photometry for follow-up and an initial investigation of their stellar properties are important scientific goals. 

So far, there have been 100-200 \mbox{$z\gtrsim8$} candidates identified in \textit{Hubble Space Telescope} (\textit{\textit{HST}}) surveys that utilize gravitational lensing by massive galaxy clusters %(e.g., \citealp{bradley14}), such as Cluster Lensing and Supernovae Survey with Hubble, (CLASH, \citealp{bradley14}) and Hubble Frontier Fields (HFF), 
and in blank field surveys (e.g., \citealp{bradley14,bouwens2015a,fink15,oesch15,ish18,morishita2018,bouwens2019,debarros2019}). %(e.g., \cite{grogin11,koekemoer11}) like The Cosmic Assembly Near-IR Deep Extragalactic Legacy Survey (CANDELS, \citealp{grogin11,koekemoer11}) and Brightest of Reionizing Galaxies/Hubble Infrared Pure Parallel Imaging Extragalactic Survey (BoRG/HIPPIES, \citealp{trenti11,yan11}). 
Photometric redshifts of this sample are largely based on rest-frame UV + optical photometry (\textit{\textit{HST}} + \textit{Spitzer}/IRAC), and only a small subset are spectroscopically confirmed. Without a spectroscopic confirmation, \textit{Spitzer} fluxes can aid in removing low-redshift interlopers from these samples. Even with a spectroscopic confirmation, \emph{Spitzer}/IRAC (rest-frame optical) fluxes are essential for robust measurements of stellar properties \citep{gonzalez2011,rya14,salmon15}.%, since at \mbox{$z\gtrsim6$} the rest-frame 4000\si{\angstrom} break is shifted to \mbox{$\gtrsim3.6\mu m$} \citep{rya14}. 

Here we use \textit{HST} and \emph{Spitzer}/IRAC imaging data from the Reionization Lensing Cluster Survey (RELICS, PI Coe) and companion survey, \emph{Spitzer}-RELICS (S-RELICS, PI Brada{\v c}) to probe rest frame optical wavelengths of seven \mbox{$z\gtrsim8$} candidates originally selected with \textit{HST}. Details of the \textit{HST}-selected high-$z$ candidates can be found in \cite{salm17,salm18} (hereafter S17, S18).  We present measurements of stellar mass, SFR, and age inferred from \textit{HST} and \textit{Spitzer} broadband fluxes.

In \S \ref{obsphot} we describe \textit{HST} and \textit{Spitzer} imaging data and photometry. In \S \ref{lens} we discuss the lens models used in our analysis. In \S \ref{sedfit} we describe our photometric redshift procedure, SED modeling procedure and calculation of stellar properties. We present our SED fitting and stellar properties results in \S \ref{results} and we conclude in \S \ref{concls}. Throughout the paper, we give magnitudes in the AB system \citep{oke74}, and we assume a $\Lambda$CDM cosmology with \mbox{$h = 0.7$}, \mbox{$\Omega_m = 0.3$}, and \mbox{$\Omega_{\rm \Lambda} = 0.7$}.

\section{Observations and Photometry}\label{obsphot}

\begin{figure*}[h!!!]
\centering
    \begin{subfigure}
        \centering
        \includegraphics[width=14cm]{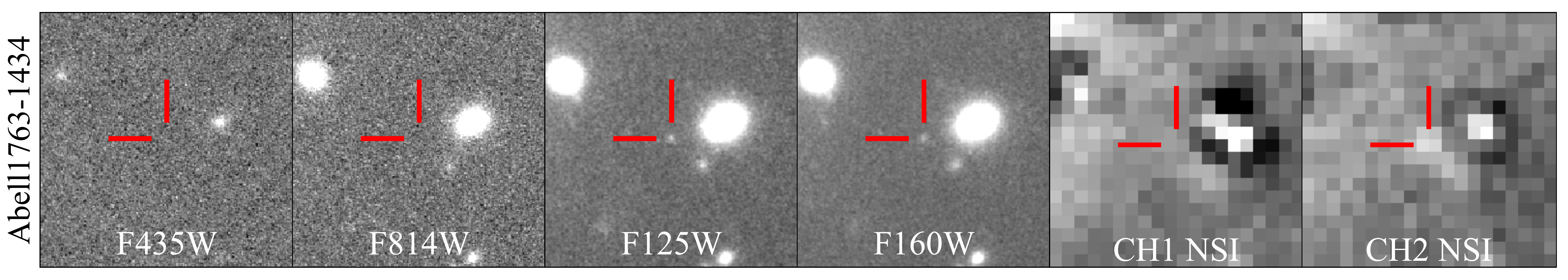}
    \end{subfigure}
    \begin{subfigure}
        \centering
        \includegraphics[width=14cm]{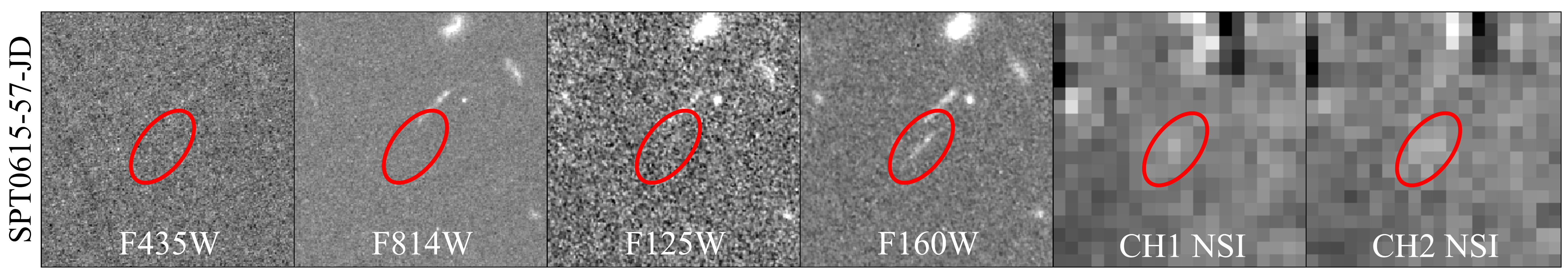}
    \end{subfigure}
    \begin{subfigure}
        \centering
        \includegraphics[width=14cm]{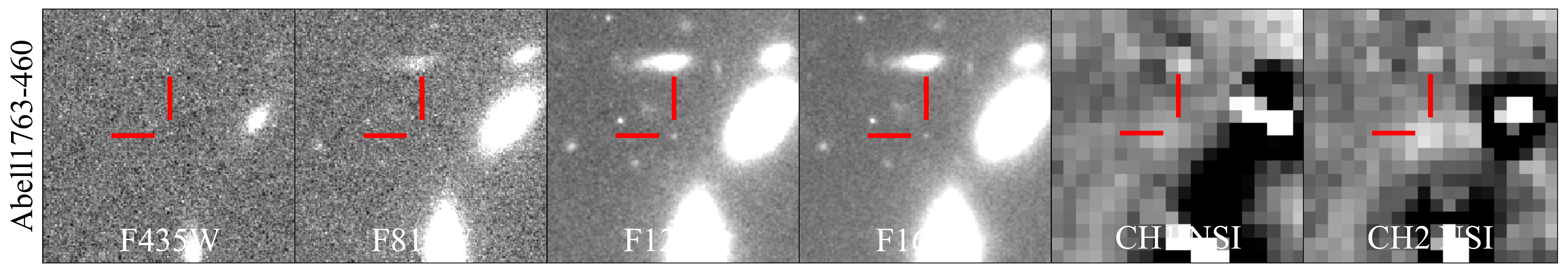}
    \end{subfigure}
    %\begin{subfigure}
    %    \centering
    %    \includegraphics[width=14cm]{abell1763-%817_cutouts.pdf}
    %\end{subfigure}
    \begin{subfigure}
        \centering
        \includegraphics[width=14cm]{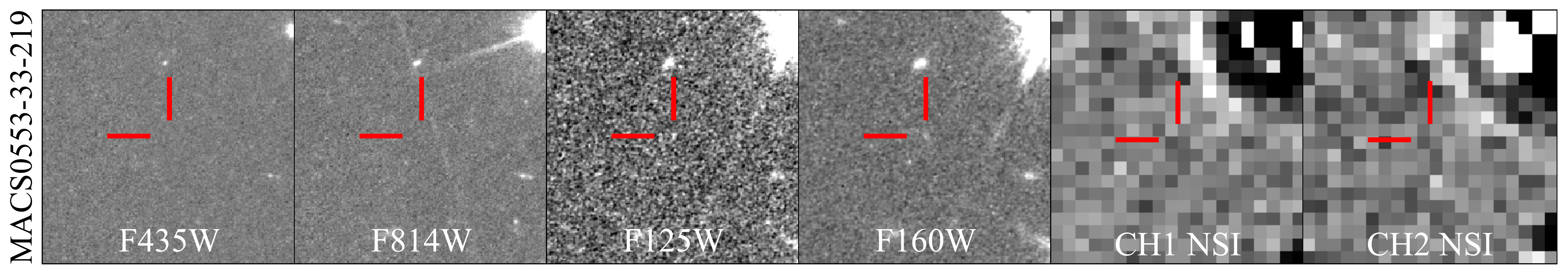}
    \end{subfigure}
    \begin{subfigure}
        \centering
        \includegraphics[width=14cm]{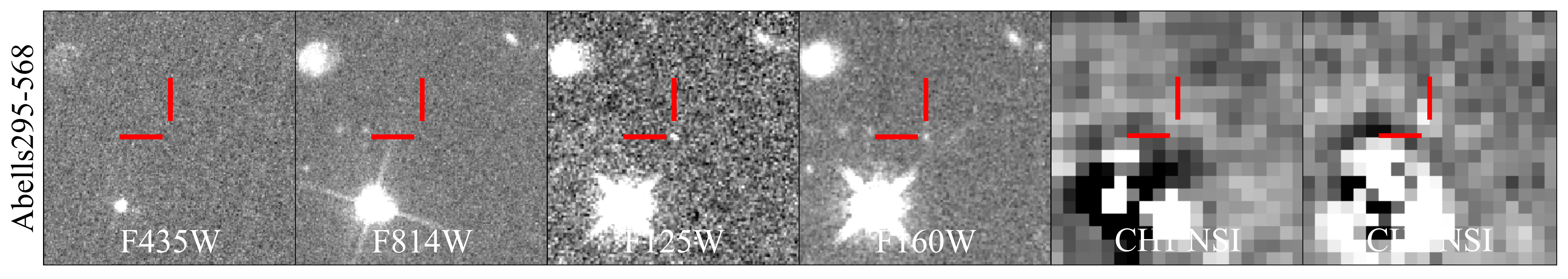}
    \end{subfigure}
    \begin{subfigure}
        \centering
        \includegraphics[width=14cm]{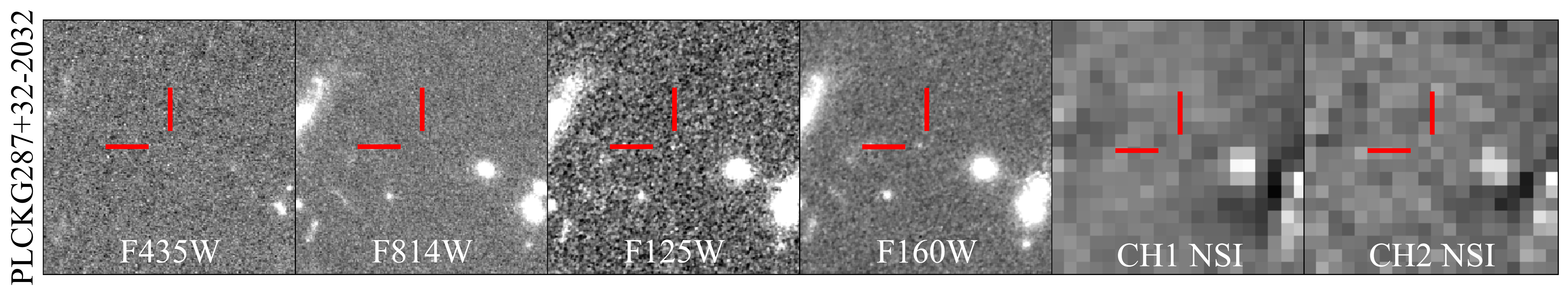}
    \end{subfigure}
    \begin{subfigure}
        \centering
        \includegraphics[width=14cm]{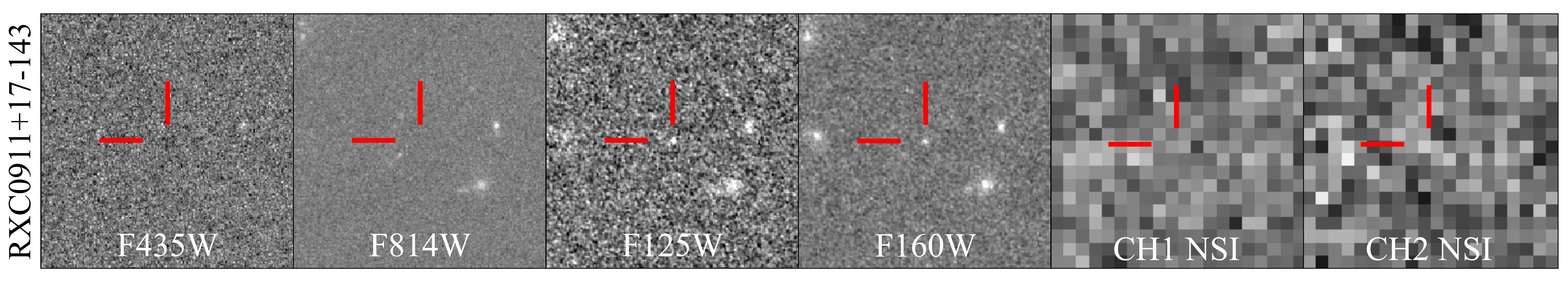}
    \end{subfigure}
\caption{Image stamps for each candidate, $12 \arcsec \times 12\arcsec$, of two ACS bands (F435W and F814W), two WFC3 bands (F125W and F160W), \textit{Spitzer}/IRAC Ch1 ([3.6]) and Ch2 ([4.5]). The \textit{Spitzer} cutouts are neighbor-subtracted images (NSI), i.e., everything in the field is subtracted except the high-$z$ source. Red lines mark the location of the source.}
\label{stamps}
\end{figure*}

\textit{HST} reduced images and catalogs are publicly available on Mikulski Archive for Space Telescopes (MAST\footnote{https://archive.stsci.edu/prepds/relics/}) and \textit{Spitzer} reduced images on NASA/IPAC Infrared Science Archive (IRSA\footnote{https://irsa.ipac.caltech.edu/data/SPITZER/SRELICS/}). Details of the survey can be found in \cite{coe19}. Here we focus on the six clusters with \mbox{$z\gtrsim8$} candidates (Abell 1763, MACSJ0553-33, PLCKG287+32, Abell S295,  RXC0911+17, and SPT0615-57, %). Select \textit{HST} and \textit{Spitzer}/IRAC cutouts and residuals are shown in 
Figure \ref{stamps}).

\subsection{\textit{HST}}
Each cluster was observed with two orbits of WFC3/IR imaging in F105W, F125W, F140W, and F160W and with three orbits in ACS (F435W, F606W, F814W), with the exception of Abell1763 which received seven additional WFC3/IR orbits. 
In this work, we use the catalogs based on a detection image comprised of the $0.06 \arcsec$/pix weighted stack of all WFC3/IR imaging, optimized for detecting small high-$z$ galaxies, described in \cite{coe19}. 

\begin{deluxetable*}{llllllllllllllllll}
\tabletypesize{\footnotesize}
\tablecaption{\label{tbl-1} \mbox{$z\gtrsim 8$} galaxy candidates and selected photometry} 
\tablewidth{0pt}
\tablehead{
\colhead{Object ID} &  \colhead{R.A.} & \colhead{Dec.} & \colhead{$\rm{F160W}^{\tablenotemark{1}}$}&\colhead{Ks} & \colhead{$[3.6]^{\tablenotemark{2}}$} & \colhead{$R_{3.6}^{\tablenotemark{3}}$} & \colhead{$[4.5]^{\tablenotemark{2}}$} & \colhead{$R_{4.5}^{\tablenotemark{3}}$} \\ \colhead{} & \colhead{(deg.)} & \colhead{(deg.)} & \colhead{(mag)} & \colhead{(mag)} & \colhead{(mag)}& \colhead{} & \colhead{(mag)} & \colhead{} 
}
\startdata
Abell1763-1434 &  203.8333744 & +40.9901793 & $26.1\pm 0.1$  & & $25.5\pm 0.4$ & 0.29 & $24.5\pm 0.2$ & 0.28 \\
Abell1763-0460 &  203.8249758 &  +41.0091170 & $27.8\pm 0.2$ & & $>$25.9 & 0.37 & $24.5\pm0.2$ & 0.39 \\
MACS0553-33-0219 & 88.3540349 & -33.6979484 & $27.2\pm 0.2$  & & $25.0\pm 0.3$ & 0.34 & $25.5\pm0.6$ & 0.34 \\
PLCKG287+32-2032  & 177.7225936 & -28.0850703&  $26.7\pm 0.2$ &$26.6\pm0.3$ & $>$26.6 & 0.53 & $>$26.4 & 0.59 \\
SPT0615-JD &  93.9792550 & -57.7721477 & $25.8\pm 0.1$ &  & $26.0\pm0.6$ & 1.43 & $25.4\pm0.4$ & 1.13 \\
RXC0911+17-0143 & 137.7939712 & +17.7897516 & $26.5\pm 0.1$ &&  $>$26.4 & 0.05 & $>$26.1 & 0.04 \\
AbellS295-0568 &  41.4010242 & -53.0405184 & $26.3\pm 0.1$ & &  $>$26.2 & 0.16 & $>$26.3 & 0.14 \\

\hline \\
%abell1763-1434 0.5836510781762962 0.12215988822267358 0.23044717992898442 0.10833283057901844 0.12374299999999999 0.0089984

%abell1763-460 0.5456299284789196 0.13225123469889435 -0.002377648027625338 0.16383134253997145 0.027961800000000002 0.0044329

%abell1763-817 -0.06608281250199688 0.07869978887385155 -0.07272076342814728 0.0784870004498743 0.0482027 0.0058312

%macs0553-33-219 0.230804211462592 0.17558883617302767 0.36375258027973967 0.11896074009258836 0.0482232 0.0087721

%plckg287+32-2032 -0.05329606416547725 0.10095443281869561 -0.11452747158317664 0.08293231390647243 0.0817813846263125 0.029547432726771048 0.07301579999999999 0.0122463

%spt0615-57-336 0.2589953976557203 0.11015463942155912 0.14937148078411844 0.11545881771031857 0.1824614 0.0165665

%rxc0911+17-143 0.1243337345588781 0.13265448226316812 0.036626652920058976 0.10423306236415723 0.0957955 0.012061299999999999

%abells295-568 0.07269366864533369 0.11220795164115452 -0.03673681805542889 0.12336269004642043 0.1048055 0.0124079

\tablenotetext{1}{Total lensed magnitude (\texttt{FLUX}\_{\texttt{ISO}})}
\tablenotetext{2}{\textit{Spitzer/IRAC} Channels 1 and 2 magnitudes measured with the same aperture as HST magnitudes and 1-$\sigma$ error. If detection is $<$ 1-$\sigma$, 1-$\sigma$ lower limit is reported.}
\tablenotetext{3}{Covariance index for \textit{Spitzer}/IRAC channels (Section \ref{spitzer})}

\end{deluxetable*}

\subsection{Spitzer Data and Photometry}\label{spitzer}
Each cluster was observed with \textit{Spitzer}/IRAC by a combination of RELICS programs (PI Soifer, \#12123, PI Brada{\v c} \#12005, 13165, 13210) and archival programs. PLCKG287+32, Abell 1763 and SPT0615-57 were observed for 30 hours each in [3.5] and [4.6] channels, including archival data (PI Brodwin \#80012). Abell S295 was observed for 5 hours in each channel, including archival data (PI Menanteau \#70149). MACS J0553-33 was observed for 5.2 hours in each channel including archival data (PI Egami \#90218). RXC0911+17 was observed for 5 hours with archival data only (PI Egami \#60032). In addition to \textit{Spitzer} and \textit{HST} fluxes, we include Ks imaging from VLT-HAWK-I (\#0102.A-0619, PI Nonino) for PLCKG287+32 (other clusters do not have such data at present). Reduction details for Ks imaging of will be detailed in Nonino et al. (2019, in prep.). 

\textit{Spitzer} data reduction and flux extraction is similar to that of the \textit{Spitzer} UltRa-Faint Survey Program (SURFSUP, \citealp{hua16}). Full details, including treatment of ICL, will be described in detail in an upcoming catalog paper (Strait et al., 2019 in prep).  
Due to the broad point spread function (PSF) and low resolution ($0.6 \arcsec$/pixel) of \textit{Spitzer} images, we extract fluxes using T-PHOT \citep{merl15}, designed to perform PSF-matched, prior-based, multi-wavelength photometry as described in \cite{merl15,merl16}. 
We do this by convolving a high resolution image (in this case, F160W) using a low resolution PSF transformation kernel that matches the F160W resolution to the IRAC (low-resolution) image and fitting a template to each source detected in F160W to best match the pixel values in the IRAC image. 

We assess the trustworthiness of the output fluxes  
using diagnostic outputs $R_{3.6}$ and $R_{4.5}$ (see Table \ref{tbl-1}),  
defined as the ratio between the maximum value in the covariance matrix for a given source (i.e., the covariance with the object's closest or brightest source) and the source's own flux variance. Covariance indices $R_{3.6}$ and $R_{4.5}$ are indicators of whether a source is experiencing confusion with a nearby source.
In the case of severe confusion and a high covariance index ($R_{3.6}$, $R_{4.5} > 1$), we perform a series of tests involving the input of simulated sources of varying brightnesses to test the confusion limit of that pair of sources. The only source with $R_{3.6}$, $R_{4.5} > 1$ in our sample is SPT0615-JD, and as described in S18, we find that simulated magnitudes brighter than $\sim25$ can be safely recovered, and we conclude that the 1-$\sigma$ flux limits (both $\sim25$) we extract are trustworthy as lower limits in magnitude (i.e., the flux of the source could be fainter than the extracted fluxes but not brighter). 

\subsection{Sample Selection}
The selection criteria of all high-redshift ($z_{\rm{peak}}\geq5.5$) \textit{HST}-selected RELICS objects is described in S17 (for details on how $z_{\rm{peak}}$ was calculated, see \S\ref{calcs}). 
This paper focuses on \mbox{$z \sim 8$} objects from the S17 sample and the \mbox{$z \sim 10 $} object from S18, that, when \textit{Spitzer} fluxes are included in their photometry, still have $z_{\rm{peak}} \geq 7.5$. 
Of the eight \mbox{$ z \sim 8 $} candidates in S17, we find that six remain likely to be at $z\geq7.5$ upon inclusion of \emph{Spitzer} fluxes (Table \ref{tbl-1}). 
The other two \mbox{$z\sim8$} candidates from S17 (SPT0615-57-1048 and PLCKG287+32-2013) were %found to have peak $P(z)$s of \mbox{$z=7.20$}, \mbox{$z=7.24$} and \mbox{$z=7.41$}, respectively, and were
moved into the \mbox{$z\sim7$} bin. We will explore these candidates in a future work.

\section{Lens Models}\label{lens}
In order to correct for magnification from lensing, relevant for SFRs and stellar masses, we use lens models created by the RELICS team. We use three lens modeling codes to produce the models for the clusters described here: \texttt{Lenstool} \citep{jullo2009} for MACS0553-33 and SPT0615-57, Glafic \citep{ogu10} for RXC0911+17, and a light-traces-mass method (LTM, \citealp{zitrin13}) for Abell S295, PLCKG287+32, and Abell 1763. Full details of the SPT0615-57 \texttt{Lenstool} model can be found in \cite{paternomahler18}, and the LTM models for Abell S295 and PLCKG287+32 are described in detail by \cite{cibirka18} and \cite{zitrin17}, respectively.

The remaining three clusters will have details available in the future, and all models are available on MAST\footnote{https://archive.stsci.edu/prepds/relics/}. Our \texttt{Lenstool} model of MACS0553-33 uses nine multiply-imaged systems as constraints, three which are spectroscopically confirmed, and our Glafic model of RXC0911+17 uses three multiply imaged systems with photometric redshifts as constraints.

We find no clear multiple image constraints in Abell 1763, but are able to make an approximate model for the cluster using the LTM method, which relies on the distribution and brightness of cluster galaxies. One should be cautious in interpreting magnifications in this field, however, because in the case of Abell 1763 where there are no visible constraints, we adopt a mass-to-light normalization using typical values from other clusters.  
Median magnifications for the high-$z$ candidates are listed in Table \ref{tbl-2}, and treatment of their statistical uncertainties is described in \S \ref{mcsim}.

\begin{figure*}[h!!!]
    \begin{subfigure}
        \centering
        \includegraphics[width=9cm]{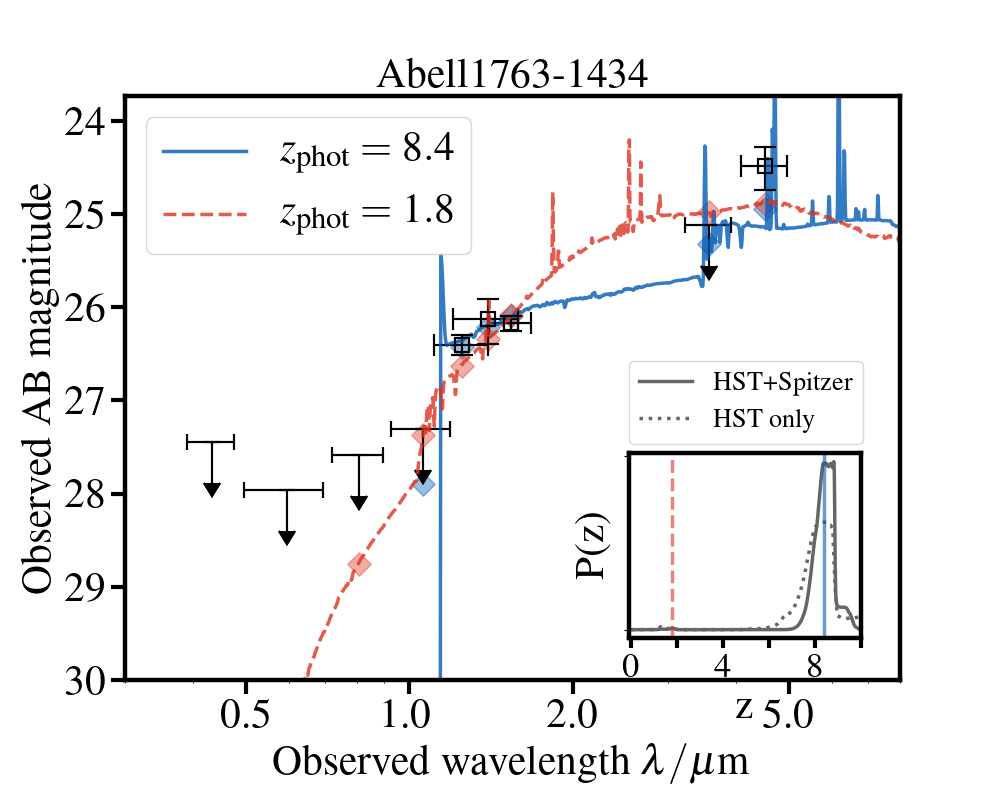}
    \end{subfigure}
    \begin{subfigure}
        \centering
        \includegraphics[width=9cm]{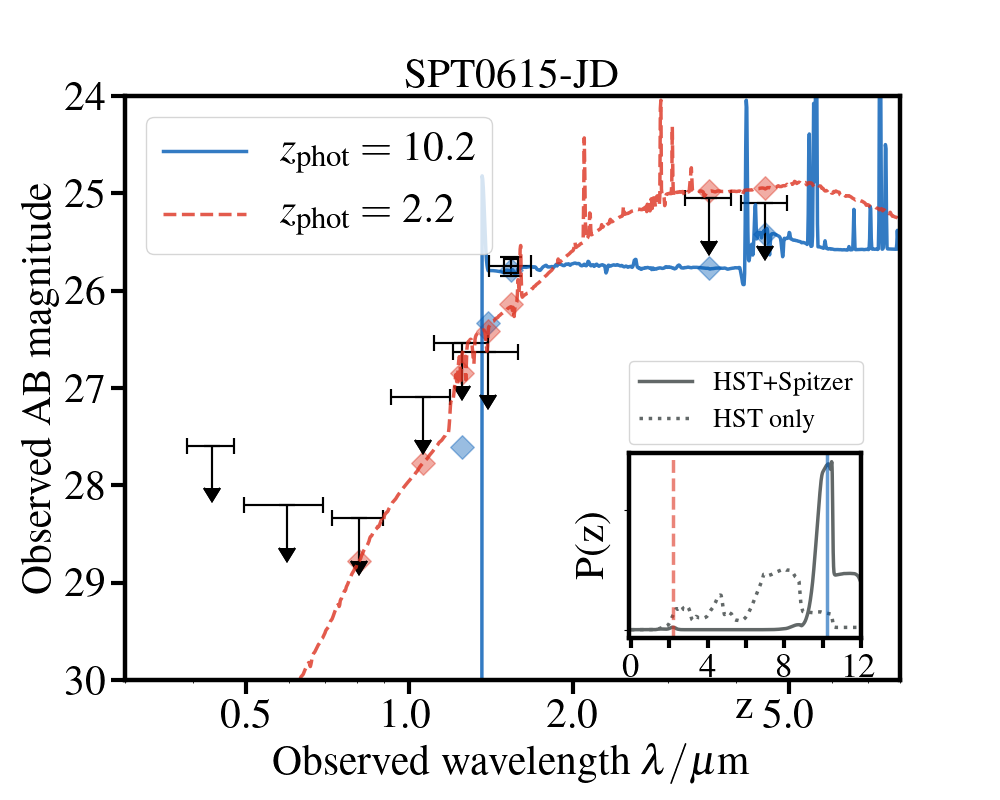}
    \end{subfigure}

\caption{Best fit SEDs for Abell1763-1434 and SPT0615-JD, fit to BC03 templates assuming a constant star formation history (CSF), 0.02\mbox{$Z_{\odot}$} metallicity (m32), Lyman-$\alpha$ escape fraction \mbox{$f_{\rm esc}=20\%$}, and small magellenic cloud dust law (SMC). Solid blue lines show best fit templates and dashed red lines show templates best fit at the associated low redshift peak in $P(z)$. Translucent blue diamonds show expected photometry for best fit and translucent red diamonds show expected photometry for low redshift fit. Inset: $P(z)$ calculated from EA$z$Y while allowing for linear combinations of default base set of BC03 templates is shown. Solid gray line shows probability with HST and \textit{Spitzer} fluxes, dotted gray shows probability with HST only fluxes. Vertical lines correspond to best fit and low redshift best fit solutions.}
\label{seds}
\end{figure*}

\begin{figure*}[h!!!!]
\centering
    \begin{subfigure}
        \centering
        \includegraphics[width=18cm]{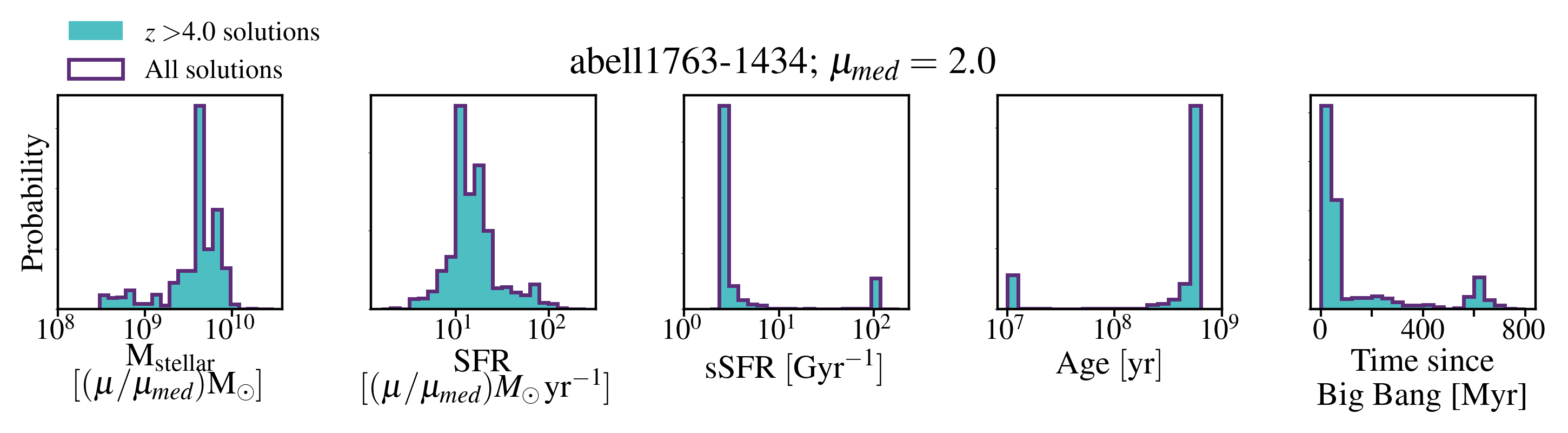}
    \end{subfigure}
    \begin{subfigure}
        \centering
        \includegraphics[width=18cm]{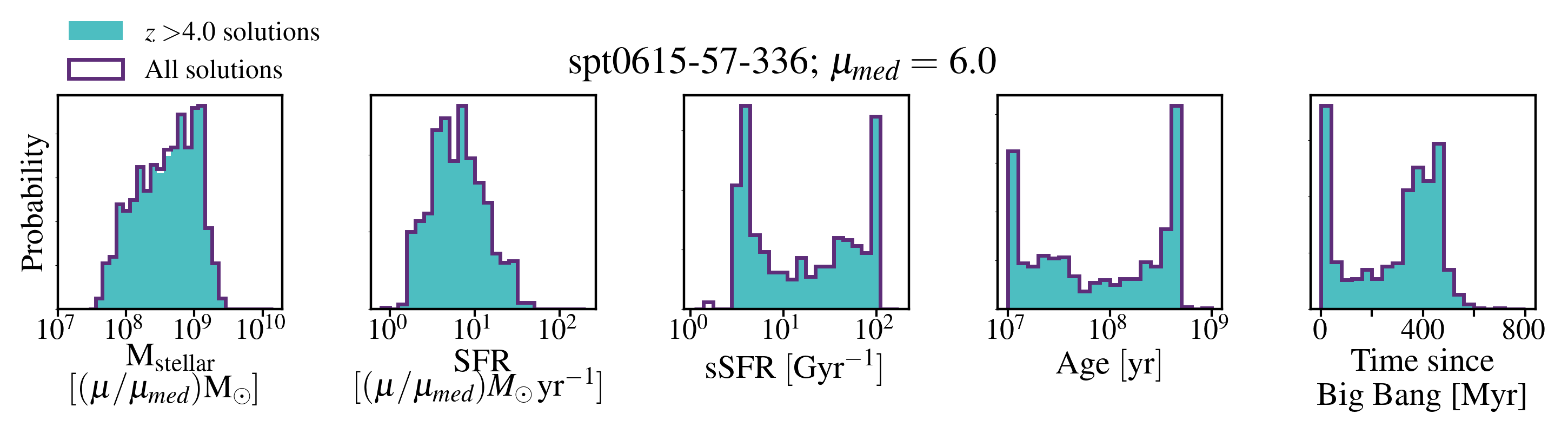}
    \end{subfigure}
\caption{Distributions of stellar properties for Abell1763-1434 (top) and SPT0615-JD (bottom) explored by Monte Carlo simulation described in \S \ref{mcsim}. From left to right, each panel shows stellar mass, star formation rate, specific star formation rate (SFR/$M_{\rm stellar}$), age, and time since the Big Bang until the onset of star formation. High-redshift solutions are shown in turquoise and all solutions, including low redshift ones, are shown in purple outline. There is no significant distinction between the two as probability for low redshift is small. The redshifts explored by the MC simulation reflect the shapes of respective PDFs for each object in Figure \ref{seds}. }
\label{mcmc}
\end{figure*}

\section{SED Fitting}\label{sedfit}

\subsection{Photometric Redshifts and Stellar Properties}\label{calcs}
To obtain a probability distribution function (PDF) and peak redshift, we use Easy and Accurate Redshifts from Yale (EA$z$Y, \citealp{bram08}), a redshift estimation code that compares the observed SEDs to a set of stellar population templates. For redshift fitting, we use the base set of seven templates from (\citealp{bruz03}, BC03), allowing linear combinations. EA$z$Y performs a \mbox{$\chi^2$} minimization on a redshift grid, which we define to range from \mbox{$z=0.01-12$} in linear steps of \mbox{$\Delta z=0.01$}, and computes a PDF from the minimized \mbox{$\chi^2$} values, where we assume a flat prior. We adopt $z_{\rm{peak}}$ as our redshift measure, which is the maximum value of the P($z$).

To calculate stellar properties of our candidates, we use a set of $\sim$2000 stellar population synthesis templates, also from BC03, this time not allowing linear combinations. We assume a Chabrier initial mass function \citep{chabrier03} between 0.1 and 100\mbox{$M_{\odot}$}, metallicity of \mbox{$0.02 Z_{\odot}$}, and constant star formation history. We allow age to range from 10 Myr to the age of the universe at the redshift of the source. We assume the Small Magellanic Cloud dust law with \mbox{$E_{\star}(B-V)$}=\mbox{$E_{\rm{gas}}(B-V)$} with step sizes of \mbox{$\Delta E(B-V)=0.05$} for \mbox{$E(B-V)=0-0.5$} mag and \mbox{$\Delta E(B-V)=0.1$} for \mbox{$E(B-V)=0.5-1$} mag.

Since it has been shown that nebular emission can contribute significant flux to broadband photometry (e.g., \citealp{schaerer10,smit14}), we add nebular emission lines and continuum to the BC03 templates using strengths determined by nebular line ratios in \cite{anders03} for a metallicity of 0.02$Z_{\odot}$. In addition, we include Ly-$\alpha$, with expected strengths calculated using the ratio of H-$\alpha$ to Ly-$\alpha$ photons calculated for Case B recombination (high optical depth, $\tau \sim 10^4$) in \cite{brocklehurst71}, assuming a Ly-$\alpha$ escape fraction of 20\%. While this is, perhaps, an overestimate at these redshifts (e.g., \citealp{hayes10}, though see \citealp{oesch15} and \citealp{stark2017}), we conservatively adopt this value to allow for a higher contribution from the Ly-$\alpha$ line.

\subsection{Biases and Systematic Uncertainties}\label{bias}

Star formation history and initial mass function are known to introduce large systematic biases in age and SFR \citep{lee09}, although at high-$z$ this is alleviated to some degree due to the fact that at $z\sim8$, the universe is only $\sim$750 Myr old \citep{pacifici2016}. 
There is also a well-known degeneracy between dust, age, and metallicity parameters, so a lack of constraints on dust attenuation can lead to a large uncertainty ($\sim$0.5-1 dex) on SFR and stellar mass \citep{hua16}. This is a particularly difficult degeneracy to break for objects at $z\sim8$ because the SED near the UV slope is not well-sampled. We explore a subset of these biases in our own sample, largely finding what is reflected in the literature. We find that changing the assumed dust attenuation law from one with the shape of the SMC or Milky Way extinction law biases stellar masses and SFRs higher by \mbox{$\sim 0.5$} dex. On the other hand, large changes in the metallicity (0.02\mbox{$Z_{\odot}$} - \mbox{$Z_{\odot}$}) introduce subdominant systematic errors on SFR, stellar mass, or age ($\lesssim$ 0.1 dex). 

An additional uncertainty is the equivalent width distribution of nebular emission lines at high-$z$, particularly [OIII] + H$\beta$, which falls in \textit{Spitzer}/IRAC [4.5] at \mbox{$z\sim8$}. Strong emission lines ($\gtrsim$ 1000\si{\angstrom}) have the potential to boost broadband fluxes, as much or more than a strong Balmer break can boost the flux, potentially biasing stellar mass, sSFR and age \citep{labbe13}. While we do not fully explore the effects of this degeneracy, we do adopt standard assumptions with regards to emission lines (\S \ref{calcs}). 
\begin{deluxetable*}{lccccccclccccccc}
\tabletypesize{\footnotesize}
\tablecaption{\label{tbl-2} Photometric Redshift and Stellar Population Modeling Results} 
\tablewidth{0pt}
\tablehead{
\colhead{Object ID}  &\colhead{$z_{\rm{peak}}^{\tablenotemark{1}}$} & \colhead{$\mu_{\rm{med}}^{\tablenotemark{2}}$}  & \colhead{$M_{\rm{stellar}} \times f_{\mu}^{\tablenotemark{3}}$} & \colhead{$SFR\times f_{\mu}^{\tablenotemark{3}}$} & \colhead{$\rm{Age}^{\tablenotemark{4}}$} & \colhead{$\rm{sSFR}^{\tablenotemark{5}}$} & \colhead{$\rm{E(B-V)}^{\tablenotemark{6}}$}  \\ \colhead{} & \colhead{} & \colhead{} & \colhead{($10^9M_{\odot}$)} & \colhead{($M_{\odot} yr^{-1}$)} & \colhead{(Myr)} & \colhead{($\rm{Gyr}^{-1}$)} & \colhead{(mag)}  
}
\startdata
%Abell1763-1434 & $8.22\pm^{0.48}_{0.46}$ & 1.6--2.5 & %$11.35\pm^{3.76}_{3.61}$ & $33.12\pm^{13.42}_{9.51}$ & %$571\pm^{70}_{167}$ & $2.60\pm^{0.96}_{0.26}$ & %$0.15\pm^{0.00}_{0.05}$   \\
%Abell1763-0817 & $7.11\pm^{0.60}_{0.75}$ & 1.3--1.7 & %$0.10\pm^{0.04}_{0.06}$ & $4.21\pm^{3.31}_{0.87}$ & %$26\pm^{11}_{13}$ & $42.96\pm^{34.86}_{11.67}$ & %$0.00\pm^{0.10}_{0.00}$ \\
%Abell1763-0460 & $8.22\pm^{0.78}_{6.50}$ & 1.9--3.8 & %$0.09\pm^{0.70}_{0.05}$ & $3.29\pm^{1.92}_{1.79}$ & %$32\pm^{609}_{22}$ & $35.88\pm^{69.08}_{35.54}$ & %$0.00\pm^{0.30}-0$ \\
%MACS0553-33-219& $7.11\pm^{1.22}_{2.27}$ & $6.7\pm$ & %$2.17\pm^{4.04}_{1.78}$ & $8.50\pm^{13.13}_{5.66}$ & %$571\pm^{235}_{516}$ & $2.60\pm^{19.41}_{0.70}$ & %$0.15\pm^{0.21}_{0.15}$   \\
%PLCKG287+32-2032  & $6.86\pm^{5.39}_{0.92}$ & $66.7\pm$ & %$0.21\pm^{0.14}_{0.08}$ & $4.68\pm^{10.30}_{4.48}$ & %$31\pm^{1403}_{17}$ & $39.94\pm^{40.88}_{35.82}$ %&$0.10\pm^{0.25}_{0.05}$  \\
%SPT0615-57-JD1 & $10.34\pm^{1.03}_{0.55}$ & $7\pm$ & %$2.58\pm^{4.18}_{1.83}$ & $38.63\pm^{46.28}_{18.36}$ & %$64\pm^{340}_{51}$ & $19.02\pm^{65.73}_{15.46}$ & %$0.05\pm^{0.05}_{0.05}$ \\
%RXC0911+17-0143 & $8.11\pm^{0.31}_{0.61}$ & $1.51\pm0.24$ & %$0.26\pm^{1.38}_{0.10}$ & $8.35\pm^{7.89}_{2.04}$ & %$34\pm^{370}_{24}$ & $33.94\pm^{71.02}_{30.37}$ & %$0.00\pm^{0.05}_{0.00}$  \\
%Abells295-0568 & $7.82\pm^{0.50}_{0.60}$ & $\mu$ & %$4.63\pm^{2.99}_{2.08}$ & $12.80\pm^{8.13}_{5.71}$ & %$571\pm^0_{70}$ & $2.60\pm^0_{0.26}$ & %$0.10\pm^{0.50}_{0.10}$ \\
%\hline \

Abell1763-1434 & $8.2^{+0.6}_{-0.2}$ & $2.0^{+0.5}_{-0.4}$ &
%$8.63\pm^{4.76}_{4.22}$--> stellar mass uncorrected
$4.3^{+2.4}_{-2.1}$& %$27.48\pm^{18.17}_{8.01}$  --> sfr uncorrected
$13.7^{+9.1}_{-4.0}$&
$510^{+60}_{-190}$ & $2.9^{+1.5}_{-0.3}$ &
$0.10\pm^{0.05}_{0.00}$   \\
%Abell1763-0817 & $7.28\pm^{0.75}_{0.66}$ & %$1.5\pm0.2$ &
%%$0.10\pm^{0.05}_{0.03}$ --> stellar mass %uncorrected
%$0.07\pm^{0.03}_{0.02}$& %$4.10\pm^{2.91}_{0.68}$ %--> sfr uncorrected
%$2.73\pm^{1.94}_{0.45}$&
%$28\pm^{8}_{11}$ & $41.18\pm^{24.46}_{8.97}$ &
%$0.00\pm^{0.50}_{0.00}$ \\
$\rm{Abell1763-0460}^{\tablenotemark{*}}$ & $8.9^{+0.2}_{-0.8}$ & $2.9^{+0.9}_{-1.0}$&
%$0.75\pm^{12.48}_{0.70}$ --> stellar mass uncorrected
$0.3^{+4.3}_{-0.2}$& %$3.55\pm^{30.27}_{1.60}$ --> sfr uncorrected
$1.2^{+10.4}_{-0.6}$&
$510^{+60}_{-500}$ & $2.9^{+102.1}_{-0.3}$ &
$0.00^{+0.25}_{-0.00}$ \\
MACS0553-33-219& $8.2^{+0.2}_{-2.4}$ & $6.5^{+0.7}_{-0.6}$ &
%$3.99\pm^{5.72}_{3.04}$ --> stellar mass uncorrected
$0.6^{+0.9}_{-0.5}$& %$11.81\pm^{16.41}_{8.46}$  --> sfr uncorrected
$1.8^{+2.5}_{-1.3}$&
$570^{+240}_{-320}$ & $2.6^{+2.9}_{-0.7}$ &
$0.20^{+0.10}_{-0.20}$   \\
PLCKG287+32-2032  & $7.9^{+0.5}_{-0.9}$ & $36^{+12}_{-6}$ &
%$0.18\pm^{0.11}_{0.06}$ --> stellar mass uncorrected
$0.005^{+0.003}_{-0.002}$& %$5.97\pm^{7.40}_{2.00}$  --> sfr uncorrected
$0.2^{+2.5}_{-0.1}$&
$31^{+9}_{-11}$ & $36.9^{+18.4}_{-7.7}$
&$0.05^{+0.10}_{-0.05}$  \\
SPT0615-JD & $10.2^{+1.1}_{-0.5}$ & $6.0^{+2.7}_{-1.7}$ &
%$2.62\pm^{4.08}_{1.87}$ --> stellar mass uncorrected
$0.4^{+0.7}_{-0.3}$& %$37.24\pm^{41.63}_{19.42}$ --> sfr uncorrected
$6.2^{+6.9}_{-3.2}$&
$80^{+320}_{-70}$ & $15.4^{+73.0}_{-11.9}$ &
$0.05^{+0.05}_{-0.05}$ \\
RXC0911+17-0143 & $8.1^{+0.4}_{-0.6}$ & $1.5^{+0.2}_{-0.2}$ &
%$0.26\pm^{1.16}_{0.12}$ --> stellar mass uncorrected
$0.2^{+0.8}_{-0.1}$& %$8.43\pm^{7.39}_{2.16}$  --> sfr uncorrected
$5.6^{+10.6}_{-1.4}$&
$34^{+200}_{-20}$ & $33.9^{+71.0}_{-28.0}$ &
$0.00^{+0.05}_{-0.00}$  \\
$\rm{AbellS295-0568}^{\tablenotemark{*}}$ & $8.1^{+0.3}_{-0.7}$ & $4.0^{+1.5}_{-0.4}$ &
%$0.26\pm^{0.13}_{0.09}$ --> stellar mass uncorrected
$0.07^{+0.03}_{-0.02}$& %$9.47\pm^{8.66}_{1.59}$ --> sfr uncorrected
$2.4^{+2.2}_{-0.4}$&
$28^{+13}_{-16}$ & $41.2^{+47.3}_{-11.9}$ &
$0.00^{+0.50}_{-0.00}$ \\
\hline \

\tablenotetext{1}{Peak redshift and 68\% CL in PDF described in Section \ref{obsphot}}
\tablenotetext{2}{Median magnification factor found using corresponding lens model. $\mu_{med}$ is assumed in SFR and $M_{stellar}$ calculations.}
\tablenotetext{3}{Intrinsic stellar mass and SFR assuming $\mu=\mu_{med}$. To use a different magnification value, use $f_{\mu}\equiv \mu/\mu_{med}$}
\tablenotetext{4}{Time since the onset of star formation assuming a constant SFR}
\tablenotetext{5}{Specific SFR, sSFR $\equiv M_{stellar}$/SFR}
\tablenotetext{6}{Dust color excess of stellar emission. SMC dust law assumed.}
\tablenotetext{*}{\textit{Spitzer} fluxes not reliable, use caution interpreting stellar properties}

\end{deluxetable*}
\subsection{Statistical Uncertainties}\label{mcsim}
To understand the statistical uncertainties from photometry and redshift in the stellar properties, we perform a Monte Carlo (MC) simulation on each object. For each iteration, we sample from the redshift PDF and recompute the photometry for each band by Gaussian sampling from the estimated errors (Table \ref{tbl-1}). In the case of upper limits, we do not perturb the fluxes. For each of 1000 iterations we use EA$z$Y to find a best fit template (from the template set for stellar properties described in \S \ref{calcs}) for the photometry, fixing the redshift to that which was sampled from the PDF on each iteration. The uncertainties on stellar properties reflect only statistical uncertainties and do not include systematic uncertainties associated with choices in initial mass function, star formation history, metallicity, dust law, or the Balmer break vs. emission line degeneracy. 

Regarding the effect of magnification uncertainties on our stellar properties, since statistical uncertainties often greatly underestimate the true uncertainties in magnification due to differences in model assumptions, we choose for simplicity to not propagate these uncertainties into those of the stellar properties but rather to assume the median magnification, $\mu_{med}$. $\mu_{med}$ and 1-$\sigma$ statistical uncertainties are listed in Table \ref{tbl-2}. To use a different magnification than is listed, one can multiply the appropriate value by $f_{\mu} \equiv \mu/\mu_{med}$.

\section{Results}\label{results}
The results from SED fitting and MC simulations are listed in Table \ref{tbl-2} as the median and 1-$\sigma$ statistical uncertainty on stellar properties for all objects. 
The redshift PDFs for all sources reflect that the high-redshift solution is preferred significantly more often than the low redshift solution in each case (\textbf{$P(z<3)\ll1\%$}). 
We find a wide range of intrinsic stellar masses ($5\times10^6 M_{\odot}$ -- $4\times10^9$ $M_{\odot}$), star formation rates (0.2-14 $M_{\odot}\rm yr^{-1}$), and ages (30-600 Myr) among the sample, and highlight, in particular, two objects showing a preference for an evolved stellar population, Abell1763-1434 and SPT0615-JD.

\subsection{Abell1763-1434}
The SED fitting and MC simulation results are shown in Figures \ref{seds} and \ref{mcmc}. We find that Abell1763-1434 is a relatively massive galaxy with an evolved stellar population. Using the assumptions outlined in \S \ref{calcs}, we find a median intrinsic stellar mass of $4.3 ^{+2.4}_{-2.1} \times 10^9 M_{\odot}$ and median age of $510^{+60}_{-190}$ Myr. The distribution of the time since the Big Bang until the onset of star formation in that galaxy is shown in the rightmost panel of Figure \ref{mcmc}, and implies that the oldest stars in this galaxy started forming $<100$ Myr after the Big Bang. 
Abell1763-1434 prefers the oldest possible solution the large majority of the time: 73\% of solutions prefer first star formation $<$100 Myr after the Big Bang, though we cannot exclude the possibility that abnormally strong nebular emission, large dust content, or some combination thereof could serve to decrease the estimated age.

We detect this source in \textit{Spitzer}/IRAC [3.6] and [4.5] at 2-$\sigma$ and 5-$\sigma$, respectively. In [4.5], the detection is significantly discrepant with the predicted photometry ($\sim$2-$\sigma$) for the high-$z$ solution (blue diamond in Figure \ref{seds}). This could be indicative of a second, younger stellar population, high levels of dust, strong [OIII] emission, or some combination thereof. Assuming the boost in [4.5] is from strong [OIII]+H$\beta$, we increase the rest-frame equivalent width from our best-fit value of $\sim215\si{\angstrom}$ to $1000\si{\angstrom}$. This exercise yields a 0.54 magnitude boost in [4.5], roughly the amount needed to match the detection. Thus, even with extreme [OIII]+H-$\beta$ equivalent widths, we still require a significant Balmer break to fit the photometry well. We are not able to fully break this degeneracy with our current data, but possible improvements include sampling the UV slope with more broad/medium-band filters (e.g., \citealp{whitaker2011}) to understand dust content, a spectroscopic redshift to mitigate redshift uncertainty, and a constraint on [OIII] equivalent width, perhaps using other emission lines such as CIII] (e.g., \citealp{maseda2017,senchyna2017}). Ultimately, \textit{JWST} will allow us to measure continuum and emission lines to resolve the degeneracy.

\subsection{SPT0615-JD}
We find that SPT0615-JD is a typical galaxy with intrinsic stellar mass of $4.4^{+6.8}_{-3.1} \times 10^8 \rm M_{\odot}$, SFR of $6.2^{+6.9}_{-3.2} \rm M_{\odot}$ $\rm{yr}^{-1}$, and a bimodal age distribution preferring either the oldest age solution possible or a younger population with first star formation $\sim$400 Myr after the Big Bang. 

Assuming $z=10.2$, the IRAC bands are uncontaminated by [OIII] + H$\beta$. The Balmer break, however, still remains fairly unconstrained due to confusion-limited \textit{Spitzer} fluxes. We report a PDF with a small secondary peak, noting an insignificant probability of a low redshift solution (\textbf{$<1\%$}). This source has two marginal detections (1--2-$\sigma$) in \textit{Spitzer}/IRAC which we plot in Figure \ref{seds} as 1-$\sigma$ lower limits in magnitude. These limits are tighter by 0.2 mag in [3.6] and 0.5 mag in [4.5] compared to fluxes reported in S18, a result of deeper data from our program that became available after the S18 analysis. This increases the probability of a high-$z$ solution and strengthens the argument made in S18 that all low-$z$ solutions require brighter \textit{Spitzer} fluxes than our upper limits allow, and all high-$z$ solutions are well-fit with fluxes fainter than the limits. 

\subsection{Other sources}
For the remaining five sources, we find a range of masses, with the least massive being PLCKG287+32-2032 at an intrinsic stellar mass of  $5^{+3}_{-2} \times 10^6 \rm M_{\odot}$. 
We report 1-$\sigma$ magnitude limits for non-detections in \textit{Spitzer} for MACS0553-33-219, %Abell1763-817, 
PLCKG287+32-2032, and RXC0911+17-143 with the exception of a 3-$\sigma$ detection in [3.6] for MACS0553-33-0219 (Table \ref{tbl-1}). 

AbellS295-568 and Abell1763-460 are both likely contaminated with bright nearby sources, and their resulting stellar properties should be interpreted carefully. 
The redshift solutions for these two sources are robust to variances in \textit{Spitzer} fluxes, and even to excluding the \textit{Spitzer} fluxes entirely.  

\section{Conclusions}\label{concls}
We present SFRs, stellar masses, ages, and sSFRs for seven \mbox{$z\gtrsim8$} candidates from RELICS. All candidates have robust high-redshift solutions ($P(z>7.5)>0.95$) after the inclusion of \textit{Spitzer}/IRAC [3.6] and [4.5] fluxes and are reasonably bright ($\leq$ 27.8 magnitudes). We highlight, in particular, Abell1763-1434 which shows evidence for an evolved stellar population at a high best-fit redshift of $z=8.2^{+0.6}_{-0.2}$, implying the onset of star formation $<100$ Myr after the Big Bang. We also present a follow-up analysis of SPT0615-JD, the highest-redshift candidate from the RELICS sample at \mbox{$z=10.2^{+1.1}_{-0.5}$}, also showing some evidence for an evolved stellar population. In both cases, a younger stellar population with extreme nebular emission, large dust content, or some combination thereof, could also explain the observed fluxes. While we cannot fully disentangle the degeneracies associated with SED fitting at $z\sim8$, all candidates presented here have interesting stellar properties that would benefit from further study with \textit{JWST}. 

\section*{Acknowledgements}
Based on observations made with the NASA/ESA Hubble Space Telescope,
obtained at the Space Telescope Science Institute, which is operated
by the Association of Universities for Research in Astronomy, Inc.,
under NASA contract NAS 5-26555. Observations were also
carried out using Spitzer Space Telescope, which is operated by the
Jet Propulsion Laboratory, California Institute of Technology under a
contract with NASA.

MB and VS acknowledge support by NASA through ADAP grant 80NSSC18K0945,  NASA/HST
through  HST-GO-14096, HST-GO-13666 and two  awards issued by Spitzer/JPL/Caltech associated with SRELICS\_DEEP and SRELICS programs.

\bibliographystyle{yahapj}
\bibliography{vstrait}

\end{document}